%% file: Traffic_simulator.tex
\documentclass[conference, a4paper]{IEEEtran}

\IEEEoverridecommandlockouts
\usepackage{graphicx,color}
\usepackage{amssymb,amsmath,times,wasysym}
\usepackage{cite}
\usepackage{epsfig}
\usepackage[linesnumbered,ruled,vlined]{algorithm2e}
\usepackage{kbordermatrix}
\usepackage{comment}

\title{ Traffic Simulator for Multibeam Satellite Communication Systems 
}
\author{Hayder Al-Hraishawi, Eva Lagunas, and Symeon Chatzinotas\\
Interdisciplinary Centre for Security, Reliability and Trust (SnT), University of Luxembourg.\\
emails: \{hayder.al-hraishawi, eva.lagunas, symeon.chatzinotas\}@uni.lu
}

\begin{document}
	\maketitle
		
\begin{abstract}
Assume that a multibeam satellite communication system is designed from scratch to serve a particular area with maximal resource utilization and to satisfactorily accommodate the expected traffic demand. The main design challenge here is setting optimal system parameters such as number of serving beams, beam directions and sizes, and transmit power. This paper aims at developing a tool, multibeam satellite traffic simulator, that helps addressing these fundamental challenges, and more importantly, provides an understanding to the spatial-temporal traffic pattern of satellite networks in large-scale environments. Specifically, traffic demand distribution is investigated by processing credible datasets included three major input categories of information: (i) population distribution for broadband Fixed Satellite Services (FSS), (ii) aeronautical satellite communications, and (iii) vessel distribution for maritime services. This traffic simulator combines this three-dimensional information in addition to time, locations of terminals, and traffic demand. Moreover, realistic satellite beam patterns have been considered in this work, and thus, an algorithm has been proposed to delimit the coverage boundaries of each satellite beam, and then compute the heterogeneous traffic demand at the footprint of each beam. Furthermore, another algorithm has been developed to capture the inherent attributes of satellite channels and the effects of multibeam interference. Data-driven modeling for satellite traffic is crucial nowadays to design innovative communication systems, e.g., precoding and beam hopping, and to devise efficient resource management algorithms.
 
\end{abstract}


\section{Introduction}\label{sec:intro}\vspace{-1mm}

Satellite systems, supported by their ability to cover wide geographic regions using a minimum amount of infrastructure on the ground, are extremely appealing to satisfy data demand ubiquity and  deliver the data rates that will be requested in the future \cite{Perez2019}. Currently the field of satellite communications is witnessing a renewed and significant attention in the global telecommunications market. 
Many emerging broadband services have recently developed for future satellite communication systems, thanks to the newly technologies that empower one satellite to manage hundreds of narrower beams with smaller coverage. These new features increase the diversification of satellite services and lead to an escalating need for satellite systems \cite{Guan2019}.
Although this is a growing challenge from the traffic volume perspective, we still have a minimal understanding about the traffic characteristics experienced by satellite systems. Such lack of knowledge prevents an efficient system design and leads to a poor resource utilization \cite{Ma2017}.

Ubiquity of satellite systems can provide services to the unserved or underserved users in different geographical regions including rural/inaccessible places and urban/suburban areas that are suffering from a lack of terrestrial infrastructure or radio resources \cite{Sharma2018}.
Therefore, the need for having traffic models tailored to a wide-range of satellite traffic types and characteristics has become essential.
Besides, understanding  traffic patterns in the large-scale environments is extremely valuable for both satellite operators and service providers. 
Traffic pattern modeling is not only necessary for developing network capacity management mechanisms but it also can be utilized in designing flexible satellite systems to adapt the spatially-heterogeneous data traffic demands, which is  a crucial requirement in broadband satellite applications \cite{Roumeliotis2019}.

Furthermore, satellite communication systems have a significant potential to integrate with 5G and beyond ecosystems to provide reliable and flexible wireless services in vast areas \cite{Di2019}. 
Specifically, in satellite communications for 5G (SAT5G) project within the initiative of 5G-infrastructure Public Private Partnership (5G PPP) \cite{5GPPP}, satellite communication proposes to contribute as an additional channel for carrying traffic over the 5G backhaul. In this setting, steering traffic load through either the satellite or the terrestrial backhauls is a thought-provoking process, especially for context-aware policies or mechanisms, where a wide variety of contexts such as type of traffic, traffic load, and location of sources has to be taken into consideration in this integration \cite{Guidotti2019}. 
This anticipated requirement motivates developing a satellite traffic simulator tool in this work that will be beneficial for performance optimality.

Generally, most of the traffic analysis and evaluation efforts focus on cellular networks due to the availability of high-quality traffic measurement tools that are accurately representing the statistical characteristics of actual traffic sources. For instance, mobile traffic patterns of thousands of cellular towers in an urban environment have been modeled in \cite{Xu2017} by using  datasets collected from local commercial mobile operators in Shanghai city in China. Other studies have used cellular network traces for characterizing and modeling cellular data traffic patterns. For example, internet traffic dynamics in large cellular networks have been studied and modeled in \cite{Shafiq2011}. Similarly, mobile phone data and application traces have been utilized in \cite{Pei2014} to investigate the urban land-use based on data traffic pattern and volume. These analyses provide a comprehensive understanding of mobile data traffic patterns of large-scale networks along with urban ecology. 
Thus, traffic  modeling is an indispensable step to achieve efficient network design and capacity planning \cite{Du2018}. 

Nevertheless the satellite traffic simulator significant role in addressing key network design and dimensioning challenges, there are only few prior related research on traffic analysis \cite{Connors1999,Jin2019}, which both are investigating Low Earth Orbit (LEO) and Intermediate Circular Orbit (ICO) satellite networks. To the best of our knowledge, traffic modeling for Geostationary Earth Orbit (GEO) satellites has not been investigated yet in the open literature. Such traffic characterization imposes complicated challenges because the type of users who access GEO satellites are varying from a low traffic single home user to very high traffic Internet backbone nodes. Beyond this, different from the terrestrial networks where the base-stations are deployed in highly populated areas, satellites deployment has a geography-centric topology, where satellite resources are accessible to all users whether inside or outside  the terrestrial network coverage. Thus, satellite traffic sources diversify from typical users include stationary users or fixed-point services to airborne and maritime platforms (airplanes, ships, etc.) \cite{Di2019}. 

Additionally, analyzing the traffic logs that collected by satellite operator is prohibitively challenging for two reasons. First, the traffic experienced by thousands of satellite terminals deployed over wide areas to ensure availability and connectivity is complicated and hard to analyze because of the redundancy and conflict logs in the collected datasets. Thus, a system with ability to clean and handle the data of large-scale traffic has to be carefully designed in order to categorize traffic patterns embedded in the thousands of terminals. Second, satellite data demand is affected by many factors, such as time and positions, etc. These factors and more may compound with each other and further complicate the analysis.

This paper fills the aforementioned gap of modeling GEO satellite traffic in the literature by developing a satellite traffic simulator with channel characterization for various broadband services. Specifically, traffic demand distribution is investigated by processing credible datasets included three major input categories of information: (i) population distribution for broadband Fixed Satellite Services (FSS), (ii) aeronautical satellite communications, and (iii) vessel distribution for maritime services. 
This work is aiming at developing a tool that combines these three dimensional datasets with some practical measurements to the coverage of multibeam satellite systems in order to extract the time domain characteristics of the geographical traffic patterns. 
The ultimate purpose of our satellite traffic simulator is to create a proving ground that can be used as input for resource management algorithms and to evaluate the innovative solutions seeking to improve satellite link utilization, performance and user experience.

\noindent
\textbf{Contributions:} The main technical contribution of this work can be summarized as follows:
\begin{itemize}
	\item Develop a traffic simulator tool for satellite services based on reliable datasets that reflect the heterogeneous  spatial-temporal traffic distributions over coverage of actual satellite beam patterns.
	\item Propose a traffic identifying algorithm that defines satellite beam borders from  coverage measurements and associates every user terminal with its serving beam. 
	Thus, the geographical context of traffic experienced by different satellite serving beams can be determined.
	\item An algorithm to determine channel coefficients of user terminals is also developed to incorporate the attributes of satellite channels and inter-beam interference.
	\item The traffic simulator provides simulation-oriented models with accurate descriptions to the heterogeneous traffic sources along with their time-domain characteristics. 
\end{itemize}

The reminder of the paper is structured as follows. System model of the proposed traffic simulator  is detailed in Section \ref{Sec:system_model}. The potential applications that can beneficial from the traffic simulator are explored in Section \ref{Sec:potential_directions}. Demonstrations of various simulation results are given in Section \ref{Sec:simulation_results}. Finally, conclusions are drawn in Section \ref{sec:conclusions}.

\vspace{2mm}
\noindent
\textbf{Reproducible research:} Satellite traffic simulator outputs, i.e., sample traffic demand instances as MAT files, are made publicly available at https://github.com/hayder-hussein/Satellite-Traffic-Simulator.

\section{System Model}\label{Sec:system_model}
\begin{figure}[!t]
	\centering
	\def\svgwidth{260pt}
	\fontsize{8}{4}\selectfont
	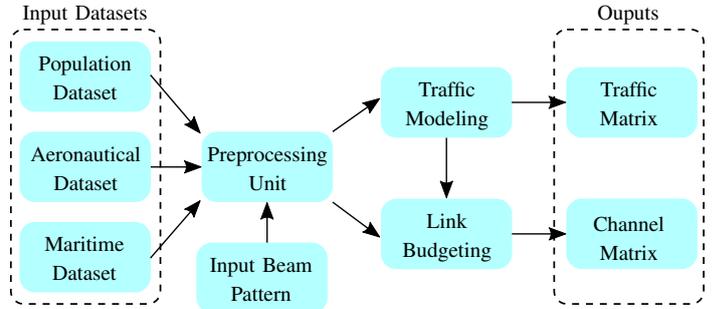 
	\caption{Block diagram of satellite traffic and system simulator.} \label{fig:system_model}
\end{figure}

Satellite traffic analysis and characterization are motivated by a key observation, that is, the traffic pattern of one beam is different from another \cite{Taricco2019}. Hence, the proposed traffic simulator analyzes and processes all of the input datasets, i.e., population, aeronautical, and maritime, in a parallel manner to model their  traffic demand and distribution within the considered beam pattern.   
The schematic diagram of the system model including the developed traffic simulator is shown in Fig. \ref{fig:system_model}, where three categories of input datasets along with the input satellite beam pattern are considered to characterize the output traffic and channel models. The functionality of each block in the diagram is elaborated in this section as follows.

\subsection{Datasets and  Beam Patterns}\label{Sec:Dataset}
In this subsection, brief descriptions of the considered datasets and the multibeam patterns are provided.

\subsubsection{Input Datasets}
The utilized datasets are obtained from reliable sources, and they are enumerated as follows: 
\begin{itemize}
	\item \textbf{Population dataset}: Population distribution for broadband FSS terminals is considered because it embeds fundamental spatial patterns of the FSS traffic. To map the traffic demand with the population data, we have downloaded the population data from  NASA Socioeconomic Data and Applications Center (SEDAC) population density database \cite{NASA}. 
	\item \textbf{Aeronautical  dataset}: The actual traffic variations in aeronautical communications is considered to study how the number of flights can affect the geographical traffic density at different time instants, and thus, the aeronautical spatial-temporal traffic distribution is accurately reflected in the simulator. To this end, the aeronautical data is extracted from anonymized and unfiltered flight-tracking source \cite{ADS_B}.
	\item \textbf{Maritime dataset}: The demand of satellite connectivity using communications on vessels is also changing remarkably with time and location. Therefore, vessel distribution for maritime is measured and evaluated by a dataset obtained from vessel traffic services (VTS), which includes ship positions and marine traffic that detected by global automatic identification system (AIS) \cite{AIS}.

\end{itemize}

\subsubsection{Input Satellite Beam Pattern}
In multibeam satellite systems, several spot beams are produced to cover a specific geographical region on Earth, and to ensure continuity, the coverage of adjacent beams are partially overlapped in very similar way to the overlay structure of cellular networks. The beam coverage patterns vary with different satellite systems and network structures, and beam gain for each satellite antenna-user pair depends on the antenna pattern and user position.
The investigated multibeam coverage pattern in this work is given in a matrix format that represents the satellite beam gain at each measured point as well as the phase rotations induced by the signal propagation. 
Specifically, a satellite beam pattern can be defined as $\mathbf{P} \in \mathbb{R}^{(\mu,4,\eta)}$, where $\mu$ accounts for the number of representative points corresponding to the sample locations where the antenna radiation pattern is given, and $\eta$ represents the number of beams. In $\mathbf{P}$, each measured point is provided in the following form $p(\text{lat},\text{long}, \gamma,\theta)$, which are latitude, longitude, channel gain in dB, channel phase, respectively.

\subsection{Preprocessing Unit}
The collected datasets need to be preprocessed because of the redundant and conflict traffic logs as well as the incomplete location information. Therefore, the preprocessing unit in the traffic simulator, shown in Fig. \ref{fig:system_model}, is responsible for eliminating these defective logs, tackling the problem of missing information, and then extracting users' positions in order to utilize them in categorizing  user terminals to their serving beams according to their geographical longitudes and latitudes. 
Additionally, the limited FSS usage in large urban areas is also taken into consideration due to the ease accessibility to  alternative broadband technologies that offered in such regions.
Since the traffic simulator is modeling the daily traffic demands on an hourly basis, the traces collected from aeronautical dataset are preprocessed by eliminating the redundant occurrences of the flights during an hour. Similarly, to capture the temporal maritime traffic demand, we analysis traces from maritime dataset and consider the position of the first occurrence of each ship during each hour within the covered area. This allows a reasonable estimate of current demand and the expected demand within an hour.
Moreover, for analyzing and modeling traffic demand per beam, the input beam pattern has to be preprocessed in order to precisely delimit the coverage of each beam. Channel  gains ($\gamma$) can be utilized to define beam borders as simple overlapped  polygons by applying a geometric triangulation algorithm that will be further explained in the next subsection.

\subsection{Per-beam Traffic Modeling}
The key of designing this traffic simulator is modeling the heterogeneous user terminals with their data demands and assorting them into the opportune serving beams according to the geographical longitudes and latitudes. Specifically, the set of operations and procedures implemented in the simulator intend to calculate the traffic at the footprint of each beam and they are outlined in  \textbf{Algorithm} \ref{alg:traffic_modeling_algorithm}. 

\begin{algorithm}[!h]
	\label{alg:traffic_modeling_algorithm}
	\SetAlgoLined
	\SetKwInOut{Input}{Input}
	\SetKwInOut{Output}{Output}
	\Input{Beam pattern $\mathbf{P}$
		\\FSS traffic vector:   $\mathcal{X}_k$,  $ k=1,\dotsc,K$
		\\Aeronautical traffic vector: $\mathcal{Y}_l$,  $ l=1,\dotsc,L$
		\\Maritime traffic vector: $\mathcal{Z}_m$,  $ m=1,\dotsc,M$
	}
\Output{Traffic matrix $[\mathbf T]$}
	\BlankLine
	\For{$ i = 1:\eta$}
	{Find $\gamma_{i_{max}}$ (maximum gain within a beam)\\
		$r=1$\\
		\For{$ j = 1:\mu$}
		{\If{$\mathbf{P}(j,3,i) >= \gamma_{i_{max}} - 3 \;\mathrm{dB}$}{
				$\mathcal{A}(r,:,i)\leftarrow\mathbf{P}(j,:,i)$
				\\ $r \leftarrow  r + 1$}
				$Lat \leftarrow \mathcal{A}(:,1,i)$, all latitudes
				\\$Long \leftarrow \mathcal{A}(:,2,i)$, all  longitudes
				\\$Z \leftarrow[Lat, Long]$
				\\$DT \leftarrow Delaunay Triangulations(Z)$
				\\$Y \leftarrow Convex Hull(DT)$ 
				\\$b_1(i) \leftarrow DT(Y,1)$, beam border latitudes 
				\\$b_2(i) \leftarrow DT(Y,2)$, beam border longitudes
				\\ $\mathcal{R}(j)\leftarrow[b_1(i),b_2(i)]$ a subset of points forms a strictly convex polygon for beam borders
		}
	\For{$\forall \mathcal{X}_k \in$ $\mathrm{polygon} \;\mathcal{R}(j)$}
		{Traffic matrix $[\mathbf T]\leftarrow \mathcal{X}_k$}
	\For{$\forall \mathcal{Y}_l \in$ $\mathrm{polygon} \;\mathcal{R}(j)$}
		{Traffic matrix $ [\mathbf T] \leftarrow \mathcal{Y}_l$}
	\For{$\forall \mathcal{Z}_m \in$ $\mathrm{polygon} \;\mathcal{R}(j)$}
	{Traffic matrix  $[\mathbf T] \leftarrow \mathcal{Z}_m$}	
	}
	\caption{Traffic Modeling Algorithm}
\end{algorithm} 

The inputs to this algorithm are the satellite beam pattern, as earlier described, and the traffic vectors of FSS, aeronautical, and maritime after preprocessing their respective datasets.
To begin with, this algorithm defines the coverage of each satellite beam through determining its borders. Footprints of beams are basically formed by a given value of the signal power on earth that must be exceeded. The signal power decreases according to the antenna characteristics when user terminal moves out of  the main lobe center of the beam antenna.  Thus, the antenna pattern with considering the view angles determines the beam's boundaries, which is usually a gain decrease of 3 dB \cite{Lutz2012}. The complete coverage of the beam pattern must be arranged in such a way that the whole footprint is filled with cells. Therefore, triangulations of a point set can be considered a useful tool in this context to delimit beam coverage. In particular, there is an interesting relationship between convex hulls and Delaunay triangulations, where a given set $P$ of points in the Euclidean space
$\mathbb{E}^m$ of dimension $m$ can lift these points onto a paraboloid  in the space $\mathbb{E}^{m+1}$ of dimension $m+1$, and that the Delaunay triangulation of $P$ is the projection of the downward-facing sides of the convex hull of the set of lifted points \cite{Gallier2013}. As a simple example, consider a point $p = (x, y)$ in the plane $\mathbb{E}^2$ is lifted to the point $l(p) = (X,Y,Z)$ on the paraboloid in $\mathbb{E}^3$, where $X = x$, $Y = y$, and $Z = x^2 + y^2$. Applying similar steps for all points in the beam pattern, we can determine sets of points represent latitudes $b_1(i)$ and longitudes $b_2(i)$ of beam borders.

Next, the traffic simulator algorithm associates every heterogeneous traffic input to its serving satellite beam by locating that user inside or on edge of the obtained polygonal region of the beam borders $\mathcal{R}(j)$. 
The three types of population-based traffic, aeronautical-based traffic, and maritime-based traffic are identified in a parallel manner to show the combination results in the traffic model in addition to their respective contribution in each beam.
For the population-based traffic, intuitively, the FSS traffic demand is proportional to population density, and thus, population distribution with a variable down-scaling factor is adjusted to map the traffic demand of FSS users.
Similarly, the proposed algorithm collects any aeronautical and maritime point lies on or inside the determined beam coverage, and then represents them in the outputs.

One of the outputs of the satellite traffic simulator is a traffic matrix $\mathbf{T}$ that contains all the identified user terminals in \textbf{Algorithm} \ref{alg:traffic_modeling_algorithm} along with their useful information, i.e., user index, beam index, location (latitude and longitude), demand type, and traffic demand, formulated altogether in a matrix form as shown in (\ref{eqn:traffic_matrix}). 

\vspace{-4mm}
\begin{eqnarray}\label{eqn:traffic_matrix}
\mathbf{T} = \!\!\!\!\! \kbordermatrix{
	& \text{User} & \text{Beam}  &\text{Lat} &\text{Long} & \text{Type}  & \text{Demand}  \\
	& u_1 		& b_i 		 &\phi_1    &\psi_1	& 1 & x_{u_1} \\
	& u_2 		& b_i 		 &\phi_2	&\psi_2	& 2 & x_{u_2} \\
	& \vdots 	& \vdots 		 &\vdots 	& \vdots& \vdots & \vdots\\
	& u_N 		& b_\eta 	 &\phi_N 	&\psi_N	& 3 & x_{u_N}
}, \!\!\!\!\!
\end{eqnarray} 

\noindent
where  $N \leq (K+L+M)$ represents the total number of the identified users. In (\ref{eqn:traffic_matrix}), each row of the traffic matrix represents a user $u_n$ that belongs to $b_i$ beam and located on latitude $\phi_n$ and longitude $\psi_n$, The tuple $\{1,2,3\}$ accounts for the traffic type and corresponds to FSS, aeronautical, and maritime, respectively, and it is represented in  the fourth  column. Finally, the traffic demand of each user is given in Mpbs on the fifth column and denoted as $x_{u_n}$.

\subsection{Link Budgeting}
The other output of the developed simulator is a per-terminal channel matrix that incorporates the inherent attributes of satellite channels and the effects of multibeam interference. Hence, channel coefficients can be obtained for every identified user in the traffic matrix by running \textbf{Algorithm \ref{alg:link_budgeting_algorithm}}. The inputs to this algorithm are (i) the obtained traffic matrix $\mathbf{T}$, and (ii) channel coefficients matrix  $\mathbf B \in \mathbb{R}^{(\mu,\eta)}$ that captures the effects of multibeam radiations and channel characteristics of the given measurements. This matrix $\mathbf B$ is extracted from the investigated satellite beam patterns by putting together the measured points with their channel coefficients ($\gamma$ and $\theta$) with respect to all satellite beams.

\vspace{-2mm}
\begin{algorithm}\label{alg:link_budgeting_algorithm}
	\SetAlgoLined
	\SetKwInOut{Input}{Input}\SetKwInOut{Output}{Output}
	\Input{Traffic matrix  $\mathbf T$
		\\Channel coefficient matrix $\mathbf B$
	}
	\Output{Channel matrix $\mathbf H$}
	\BlankLine
	\For{$ n = 1:N$}
	{	$\gamma'_n = \text{Interpolate} \; u_n$ within its beam points\\
		Find earth-satellite distance ($d_n$) of $u_n$ from (\ref{eqn:distance}) \\
		Calculate pathloss $PL_n(dB)$ from (\ref{eqn:path_loss})\\
		\For{$ j = 1:\mu$}
		{$\mathbf{d} \leftarrow \text{geo disatnces}\;(\phi_n,\psi_n) \; \text{and} \;\mathbf{B}(:,j)$
			\\$\delta_{min} \leftarrow index\left(\min (\mathbf{d}(:))\right) $ 
			\\$G_{n,j} \leftarrow 10 \log\left( \left|\mathbf{B}(\delta_{min},j)\right|^2\right)   $, gain in dB
			\\$h_{n,j} \leftarrow G_{n,j} - PL_n(dB) + G_{Rx}(dB)$
			\\$A_{n,j} \leftarrow 10^{(h_{n,j}/20)}$
			\\$a_{n,j} \leftarrow A_{n,j}\exp (2\pi i \mod(d_n,\lambda)/\lambda) $
			\\	$\mathbf{H}(n,j) \leftarrow a_{n,j} $ 
		}
	}
	\caption{Link Budgeting Algorithm}
\end{algorithm}

First, channel gain of user ($\gamma'_n$) can be computed by applying interpolation with the values of nearby cluster of sample points in the beam pattern \cite{linkbudget2019}. Next, path loss of satellite link at user $u_n$ can be calculated in dB as
\begin{eqnarray} \label{eqn:path_loss}
	PL_n = 20 \log\left( \frac{4\pi d_n}{\lambda}\right),
\end{eqnarray}
where $\lambda$ is the wavelength and $d_n$ is the distance between the $n$-th user and the satellite, which can be determined as \cite{Jiang2010}

\begin{small}\vspace{-3mm}
\begin{eqnarray} \label{eqn:distance}
	&&\!\!\!\!\!\!\!\!\!\!\!\!d_n=(R+h) \times \nonumber\\
	&&\!\!\!\!\!\!\!\!\!\!\!\!\sqrt{\! 1\!+\!\!\left(\!\frac{R}{R\!+\!h}\!\right)^{\!\!2} \!\!\!-\!\!\frac{2R}{R+h}\left( \cos(\psi_l \!-\!\psi_n) \cos\phi_l\cos\phi_n \!+\!\sin\phi_l\sin\phi_n\right) }\nonumber\\
\end{eqnarray}
\end{small}

\noindent
where $R$ is the effective radius of Earth, $h$ is satellite altitude, and $\phi_l$ and $\psi_l$ are satellite latitude and longitude, respectively. To model the impacts of multibeam satellite radiations, channel complex coefficients are calculated for each user with respect to all beams by evaluating the effect of the nearest sample points in each beam to the considered user ($u_n$). The distances between users and sample points are calculated by using the spherical law of cosines. Then, path loss is calculated as elaborated in \textbf{Algorithm} \ref{alg:link_budgeting_algorithm}, where $G_{Rx}$ is the receiver antenna gain.

The output resulting from the developed link budgeting algorithm, and consequently from the traffic simulator, is a channel matrix $\mathbf H$, that is defined in (\ref{eqn:channel_matrix}). This is a complex matrix
represents the channel gain and phase for each user in the traffic matrix.  This matrix aggregates all the received signals at every identified user including the desired signals plus the interference generated from all other beams.

\begin{eqnarray}\label{eqn:channel_matrix}
\!\!\!\!\!\!\!\mathbf H = \! \kbordermatrix{
			& b_1 	& b_2  & \cdots  & b_\eta  \\
	u_1		& a_{1,1}	& a_{1,2} & \cdots  & a_{1,\eta} \\
	u_2		& a_{2,1} 		& a_{2,2} 	& \cdots  & a_{2,\eta} \\
	\vdots	&\vdots 	& \vdots & \ddots  & \vdots \\
	u_N		& a_{N,1} & a_{N,2} & \cdots  & a_{N,\eta}
} \!\!\!\!
\end{eqnarray}

\vspace{2mm}
\noindent
where $a_{n,j}$ is a complex number represents  gain and phase of the $j$-th beam signal at the $n$-th user. 

\section{Potential Directions to Employ the Traffic Simulator}\label{Sec:potential_directions}
Beyond the essential role that traffic simulator can play in designing and dimensioning newly unreleased satellites, it can also be utilized in developing some emerging technologies that will enhance satellite communication competencies, which are including but not limited to the following list:
\begin{itemize}
	\item \textbf{Beam Hopping}: To optimally adapt the inconsistent traffic demands over time and geographical locations, beam hopping concept has been studied to allow satellite systems share their resources among multiple beams, and thus, offer higher usable throughput. Specifically, instead of static illumination for all serving beams, the satellite cycles in time through a set of coverages according to a schedule derived from the traffic demands. As a result, at a certain moment, only one coverage of the set is active with full power and bandwidth \cite{Lei2020}. 
	\item \textbf{Flexible Precoding} In the forward link of multibeam satellite systems, precoding techniques can effectively alleviate co-channel interference when aggressive frequency reuse is applied. However, satellite beams are generally affected by an uneven service demands, as there are some hot spots that required more carrier resources to avoid congestion. The flexibility of allocating satellite resources based on demand enables new solutions that can cope with the problem of co-channel interference introduced by the higher frequency reuse in hot spot areas, where the traffic simulator can play a vital role in providing  insights about traffic behaviors \cite{Taricco2019}.
	\item \textbf{Carrier Aggregation}: Carrier aggregation in cellular networks has achieved a considerable enhancement in performance through maximizing spectrum utilization and satisfying high throughput demands. Thus, an interesting next step would be integrating carrier aggregation into satellite architectures in synergy to harness the multiplexing gain and achieve higher peak data rate. Traffic distribution awareness in such systems leads to develop efficient load balancing and flexible carrier allocation algorithms targeting a proportionally fair user demand satisfaction \cite{Mirza2019,Hayder2020}. 
	\item \textbf{Network Functions Virtualization (NFV)}: In networking domain, NVF as a prominent technology has the potential to dramatically redefine the substance of network infrastructure, NFV refers to the virtualization of network functions carried out by specialized hardware devices and their transformation into software-based appliances.
	NVF introduces some outstanding benefits such as sharing of resources among different NFs and users, and up- and down-scaling of resources assigned to each function. 
	Network virtualization has extended to encompass satellite systems to obtain the advantages of fast set-up time as well as resource elasticity. Thus, according to the traffic simulator outputs of customer density and demand, the satellite virtual network might request to scale up or down the resources assigned to each entity \cite{Bertaux2015}. 
\end{itemize}

\section{Simulation Results}\label{Sec:simulation_results}
This section presents the simulation results to show the outputs of the traffic simulator, where some thematic maps are provided to visualize the distribution of the heterogeneous traffic demands with their serving satellite beams. To obtain practical results and meaningful insights, we consider two different realistic satellite beam  patterns, which are both collected by the European Space Agency (ESA) and provided in the context of the funded ESA FlexPreDem project \cite{ESA_FlexPreDem}, and they consist of 71 and 100 beams covering Europe whose antenna pattern gains and channel measurements are included as well. 
Additionally, the obtained traffic and channel matrices are utilized to plot traffic demands and inter-beam interference. The simulation parameters can be found in Table I. The considered coverage area spans from latitude 25 to 80 degrees and longitude $-$40 to 50 degrees.

\begin{table}[!h]
	\centering
	\caption{Simulation Parameters}
	\begin{tabular}{l|l}
		\hline
		Key                 & Value    \\
		\hline
		Orbit				& GEO\\
		Satellite longitude & 13$^{\circ}$E\\
		Satellite altitude	& 35786 km\\
		Satellite total radiated power & 6000 W\\
		Downlink carrier frequency & 19.5 GHz\\ 
		User link bandwidth & 50 MHz \\
		Number of beams         &  71 and 100 \\
				Receiver antenna gain     & 40.7 dB       \\
		\hline
	\end{tabular}
\end{table}

\begin{figure}[!t]\centering
	\includegraphics[width=0.52\textwidth]{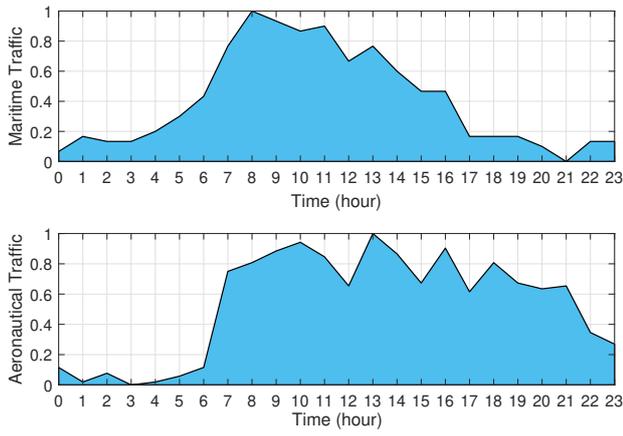}\vspace{-2mm}
	\caption{Normalized traffic profiles of maritime and aeronautical  services.}
	\label{fig:normalized_traffic_profile}\vspace{-2mm}
\end{figure}

Before exploring the data traffic modeling and delve into the demonstrations, we first visualize the temporal maritime and aeronautical traffic distributions over the geographic area located between the longitudes 10-15 degrees and the latitudes 54-58 degrees. Fig. \ref{fig:normalized_traffic_profile} shows the normalized traffic profiles of maritime and aeronautical communications during a whole day. Specifically, the top curve in Fig. \ref{fig:normalized_traffic_profile} represents maritime service traffic and we can observe that the maritime traffic demand during the day is higher than the requested traffic  during the night. Apparently, the traffic peaks are happening during the morning, which suggests that most ships tend to consume data heavily in the mornings and gradually decreases until the midnight. 
On the other hand, the aeronautical traffic demand during 24 hours of a day is analyzed, and its normalized traffic profile is depicted in the bottom curve of Fig. \ref{fig:normalized_traffic_profile}. It can be clearly seen that the aeronautical data traffic has different characteristics comparing with maritime traffic distribution, namely, the aeronautical traffic profile has several peaks within a day and remains relatively high during the night.

\begin{figure}[!t]\centering
	\includegraphics[width=0.52\textwidth]{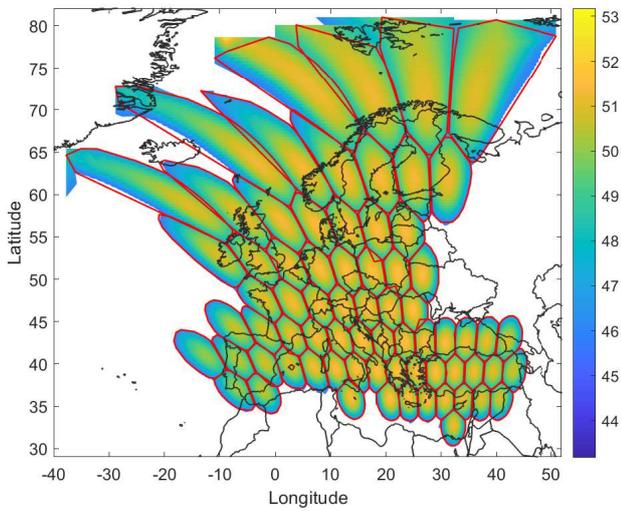}\vspace{-2mm}
	\caption{ESA 71-beam pattern covering Europe.}
	\label{fig:71_beam_pattern} \vspace{-2mm}
\end{figure}

\begin{figure}[!t]\centering
	\includegraphics[width=0.52\textwidth]{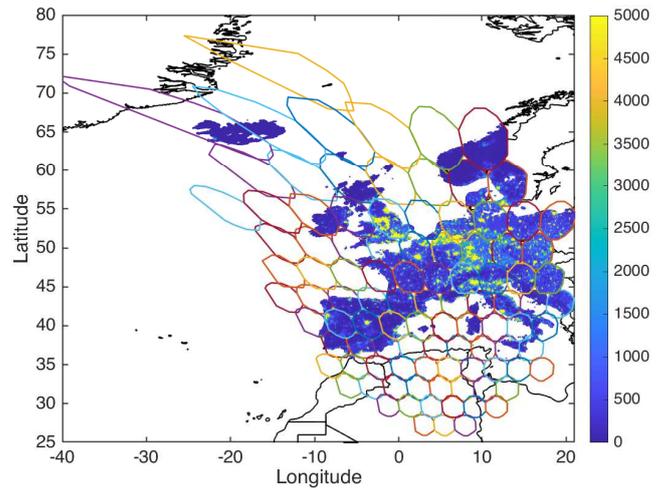}\vspace{-2mm}
	\caption{Population-based traffic model with the 100-beam pattern.}
	\label{fig:FSS_traffic}\vspace{-2mm}
\end{figure}

Next, the results of preprocessing the provided satellite beam patterns can be envisioned in Fig. \ref{fig:71_beam_pattern}, where the satellite antenna gains and beam borders of the 71-beam pattern are illustrated. 
In the following, sample results of the developed satellite traffic simulator are presented. In order to better understand the traffic patterns of the heterogeneous demands inside their respective beams, we provide visualized analysis of them over their geographical areas. The two studied beam patterns will be used interchangeably in the presentation of these results.
To begin with, the population-based model together with the considered 100-beam pattern and its obtained geographical beam borders are shown in Fig. \ref{fig:FSS_traffic}. Here, the population data has been down-scaled by a factor of 1000. Obviously, there are significant differences of data traffic across satellite beams.

\begin{figure}[!t]\centering
	\includegraphics[width=0.52\textwidth]{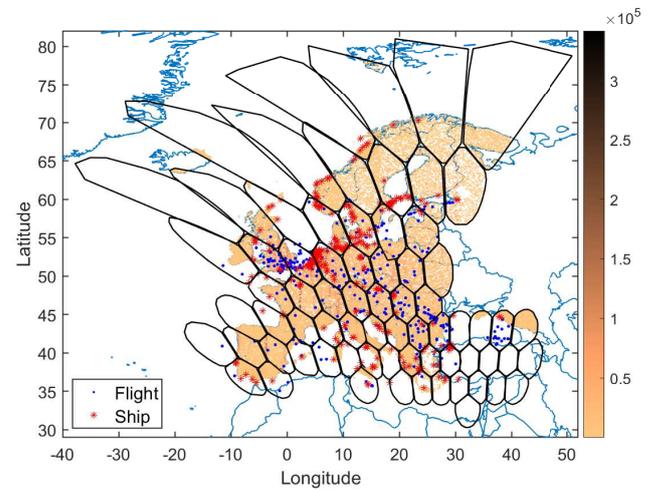}\vspace{-2mm}
	\caption{Traffic distributions over the 71-beam pattern at 4 AM.}
	\label{fig:traffic_04}\vspace{-2mm}
\end{figure}

Fig. \ref{fig:traffic_04} depicts maritime and aeronautical traffic distributions at 4 AM along with the population-based traffic model and the 71-beam pattern, where each blue dot at the map represents a flight and every red star accounts for a cruise ship. It can be noticed that most of satellite beams experience low traffic in early morning. This finding is in accordance with the observation of Fig. \ref{fig:normalized_traffic_profile}, as the traffic demand in the morning, specifically between midnight and 4 AM, is relatively low comparing to other time instances.

\begin{figure}[!t]\centering\vspace{-2mm}
	\includegraphics[width=0.52\textwidth]{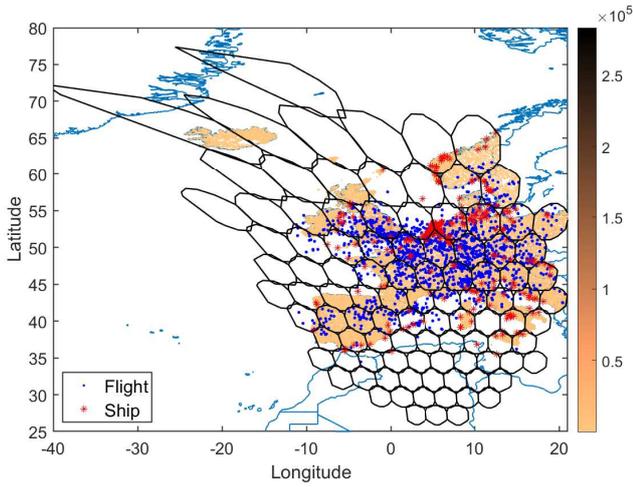}\vspace{-3mm}
	\caption{Traffic simulator output at  8 AM with the 100-beam pattern.}
	\label{fig:traffic_08}\vspace{-4mm}
\end{figure}

The heterogeneous traffic distributions over the 100-beam pattern at 8 AM are shown in Fig. \ref{fig:traffic_08}. At this time of the day, we find two observations in this investigation. First, in terms of the traffic type, it can be noticed that the aeronautical traffic demand significantly increases comparing with former time instance, and maritime data demand slightly increases. Second, data traffic volume within each distinct satellite beam varies significantly, which may lead to inefficient resource utilization and performance degradation.

\begin{figure}[!t]\centering
	\includegraphics[width=0.52\textwidth]{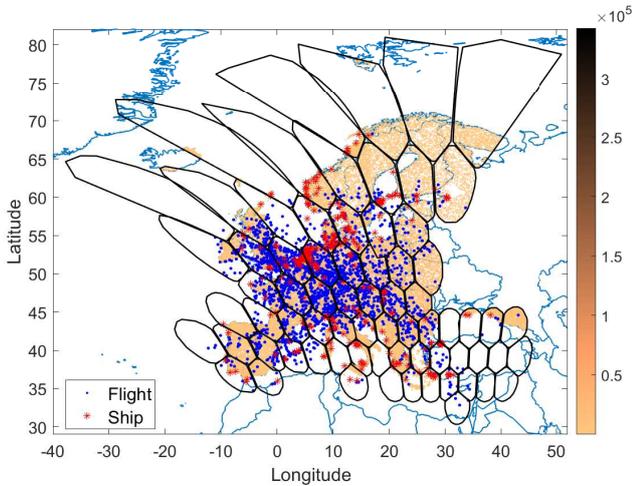}\vspace{-3mm}
	\caption{Spatial traffic distributions over the 71-beam pattern at 12 PM.}
	\label{fig:traffic_12} \vspace{-4mm}
\end{figure}

Fig. \ref{fig:traffic_12} shows traffic density for all types of data demands across the coverage of 71 satellite beams at 12 PM. Fig. \ref{fig:traffic_12} clearly reveals that the big residential areas have highly condensed movements at this time interval, which indicates that their serving satellite beams are heavily loaded with higher demand from the flights and FSS terminals. Moreover, the vessels exhibit higher density along the coasts and that means more data traffic is requested from the serving satellite beams.

 \begin{figure}[!t]\centering\vspace{-2mm}
 	\includegraphics[width=0.5\textwidth]{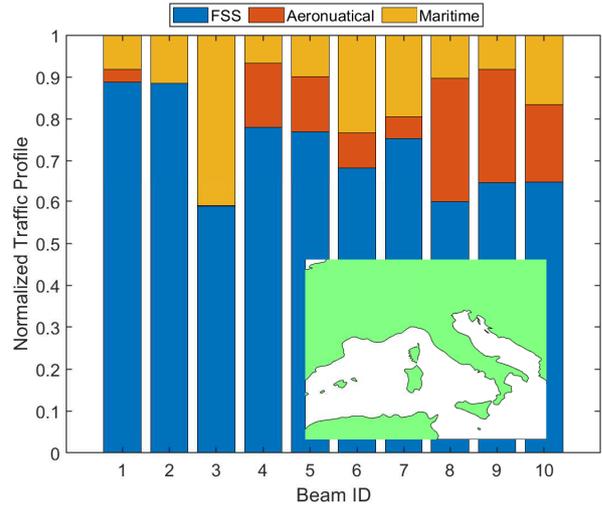}\vspace{-2mm}
 	\caption{The normalized traffic profile of the heterogeneous demands using the coverage of 71-beam pattern.}
 	\label{fig:71_beam_traffic_profile}\vspace{-4mm}
 \end{figure}
 
 \begin{figure}[!t]\centering
 	\includegraphics[width=0.5\textwidth]{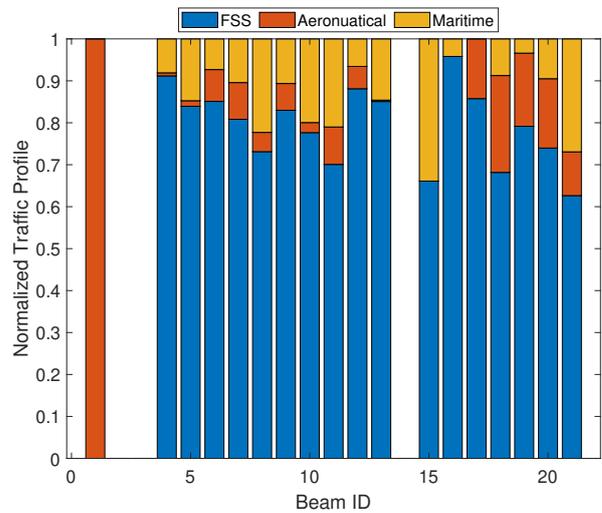}\vspace{-2mm}
 	\caption{The normalized traffic profile of the heterogeneous demands across the coverage of 100-beam pattern.}
 	\label{fig:100_beam_traffic_profile}\vspace{-4mm}
 \end{figure}

To further investigate the traffic demand behavior over various beams in different beam patterns, we have scrutinized the heterogeneous traffic demand at 4 PM over a certain geographical area that spans between 35 to 50 degrees of latitudes and  0 to 20 degrees of longitudes that shown in the map within Fig. \ref{fig:71_beam_traffic_profile}. We quantify traffic characteristic by computing the contribution ratio of each demand type, i.e. FSS, aeronautical, and maritime traffic, through analyzing and  determining their normalized traffic profiles when considering the 71 and 100 beam patterns, as presented in Fig. \ref{fig:71_beam_traffic_profile} and Fig. \ref{fig:100_beam_traffic_profile}, respectively. 
Clearly, for the same coverage area, there is a big difference in the number of serving beams between the considered beam patterns. Moreover, traffic types and loads are varying substantially from one beam to another. On one hand, the normalized traffic profile of the 71-beam pattern that presented in Fig. \ref{fig:71_beam_traffic_profile} reveals that FSS traffic has the major share, which makes sense owing to the high demand of FSS. Moreover, it can be seen that all beams are serving users and having traffic demand during the considered time interval.

On the other hand, Fig. \ref{fig:100_beam_traffic_profile} shows the normalized traffic profile of the 100-beam pattern for the same studied time iterval and geographical area.  Although the captured traffic was at 4 PM, we can see some beams are only serving flights, and also there are few beams with zero demand. This observation reveals that relying merely on the population is not a good approximation to model traffic demand, which is commonly done in the literature, e.g., \cite{Lagunas2015}. Such traffic variations are crucial to take into account in satellite resource allocation and traffic load balancing, as it directly influences the useful system capacity.
Additionally, these major deviations between the investigated beam patterns and unbalanced loads within the serving beam for the same beam patterns have to be addressed in order to improve system capacity utilization. In this context, the proposed traffic simulator can pave the way to propose more efficient resource allocation algorithms aiming at configuring  future satellite systems in a sense of providing the required capacity in a proportionally fair manner based on beam demand.

\begin{figure}[!t]\centering
	\includegraphics[width=0.52\textwidth]{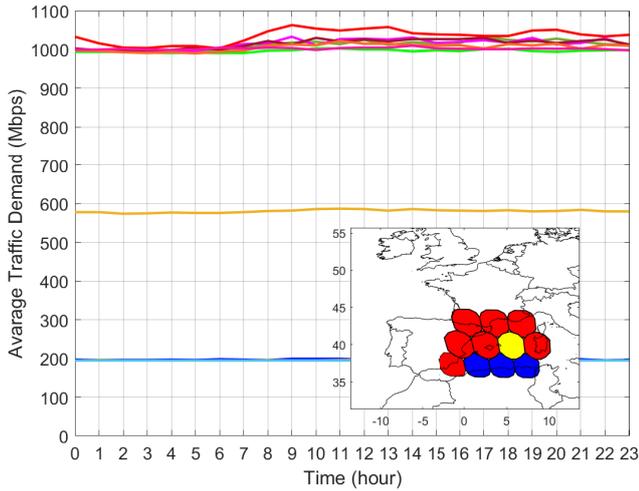}
	\caption{Average traffic demand of 11 beams per hour during a whole day.}
	\label{fig:hot_cold_beams}
\end{figure}

Fig. \ref {fig:hot_cold_beams} investigates the traffic variations among 11 beams over the area shown in the same figure, where the average traffic demand of each beam per hour during a whole day is depicted. Clearly, the demand in some beams greatly exceeds a certain system capacity (hot-spots that shown in red) while in others the situation is the opposite (cold-spots that shown in blue), and some beams have a moderate demand which can be called warm beams (shown in yellow). This deviation in demand raises a paradoxical scenario when demand is left unmet in the hot-spots while capacity is left unused in the cold-spots. Therefore, utilizing the traffic simulator in such scenarios can assist to flexibly allocate on-board resources over service coverage.

\begin{figure}[!t]\centering
	\includegraphics[width=0.5\textwidth]{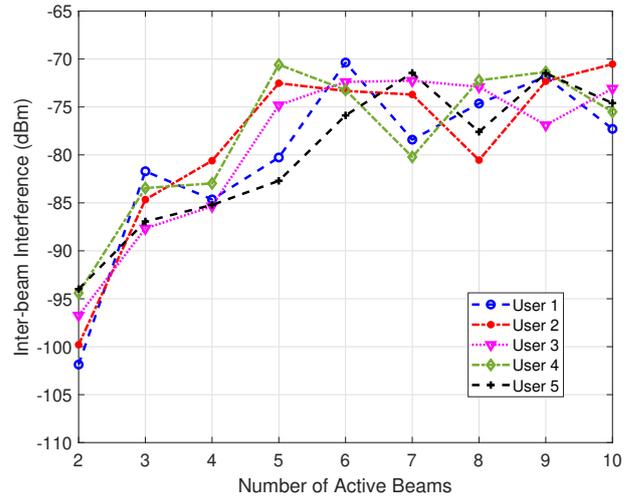}
	\caption{Inter-beam interference versus number of active beams.}
	\label{fig:Inter_beam_interference}
\end{figure}

To validate the proposed link budgeting algorithm, the inter-beam interference is investigated against the number of active beams in Fig \ref{fig:Inter_beam_interference}. Specifically, five different users have been randomly selected, and then their channel matrix is extracted by considering the 100 beam pattern and applying Algorithm \ref{alg:link_budgeting_algorithm}. Afterwards, the inter-beam interference is calculated at each user from the obtained channel matrix  by varying the number of active (transmitting) beams from 2 to 10 beams. It can be clearly seen that the inter-beam interference increases with number of active beams as it causes the lowest level of interference when there are only two active beams.

\section{Conclusions}\label{sec:conclusions}
One of the major aspects in which satellite networks differ from cellular networks, resides in large fluctuation of traffic that satellites experience in their coverage areas owing to the large diversity of traffic sources. Therefore, investigating traffic type and intensity is essential for planning and optimizing satellite systems. To this end, a satellite traffic simulator has been developed in this paper to model and typify data traffic patterns of satellite services in large-scale environments. Specifically, traffic demand distribution over Europe is investigated through processing credible datasets of three major traffic sources, and they are FSS, aeronautical, and maritime. This traffic simulator offers practical models that combines dynamic traffic usage with realistic satellite beam patterns via investigating the correlation between time-domain traffic characteristics and geographical locations of user terminals.
Our traffic simulator can play a crucial role in system design related to quality of service provisioning, as it can be readily integrated into resource allocation algorithms and other future satellite ventures. 
Moreover, the implementation of this traffic simulator offers practical and feasible steps to further extend the considered coverage area from Europe to include the entire planet, thus with the availability of appropriate comprehensive datasets we can get a worldwide satellite traffic simulator, which is our future plan and it will be released upon completion.

\section*{Acknowledgement}

This work is financially supported in part by Luxembourg National Research Fund (FNR) under the projects FlexSAT (C19/IS/13696663) and IPBG INSTRUCT.

\linespread{1.2}
\bibliographystyle{IEEEtran}
\bibliography{IEEEabrv,References}

\end{document}

%% file: system_model.pdf_tex
\begingroup%
  \makeatletter%
  \providecommand\color[2][]{%
    \errmessage{(Inkscape) Color is used for the text in Inkscape, but the package 'color.sty' is not loaded}%
    \renewcommand\color[2][]{}%
  }%
  \providecommand\transparent[1]{%
    \errmessage{(Inkscape) Transparency is used (non-zero) for the text in Inkscape, but the package 'transparent.sty' is not loaded}%
    \renewcommand\transparent[1]{}%
  }%
  \providecommand\rotatebox[2]{#2}%
  \newcommand*\fsize{\dimexpr\f@size pt\relax}%
  \newcommand*\lineheight[1]{\fontsize{\fsize}{#1\fsize}\selectfont}%
  \ifx\svgwidth\undefined%
    \setlength{\unitlength}{400.77164778bp}%
    \ifx\svgscale\undefined%
      \relax%
    \else%
      \setlength{\unitlength}{\unitlength * \real{\svgscale}}%
    \fi%
  \else%
    \setlength{\unitlength}{\svgwidth}%
  \fi%
  \global\let\svgwidth\undefined%
  \global\let\svgscale\undefined%
  \makeatother%
  \begin{picture}(1,0.44343211)%
    \lineheight{1}%
    \setlength\tabcolsep{0pt}%
    \put(0,0){\includegraphics[width=\unitlength,page=1]{system_model.pdf}}%
    \put(0.10795764,0.42582264){\color[rgb]{0,0,0}\makebox(0,0)[t]{\lineheight{1.36264527}\smash{\begin{tabular}[t]{c}Input Datasets\end{tabular}}}}%
    \put(0,0){\includegraphics[width=\unitlength,page=2]{system_model.pdf}}%
    \put(0.10898526,0.22060648){\color[rgb]{0,0,0}\makebox(0,0)[t]{\lineheight{1.36264527}\smash{\begin{tabular}[t]{c}Aeronautical\\Dataset\end{tabular}}}}%
    \put(0,0){\includegraphics[width=\unitlength,page=3]{system_model.pdf}}%
    \put(0.10703387,0.08979965){\color[rgb]{0,0,0}\makebox(0,0)[t]{\lineheight{1.36264527}\smash{\begin{tabular}[t]{c}Maritime \\Dataset\end{tabular}}}}%
    \put(0.10802633,0.35260581){\color[rgb]{0,0,0}\makebox(0,0)[t]{\lineheight{1.36264527}\smash{\begin{tabular}[t]{c}Population\\Dataset\end{tabular}}}}%
    \put(0,0){\includegraphics[width=\unitlength,page=4]{system_model.pdf}}%
    \put(0.37098008,0.22060648){\color[rgb]{0,0,0}\makebox(0,0)[t]{\lineheight{1.36264527}\smash{\begin{tabular}[t]{c}Preprocessing\\Unit\end{tabular}}}}%
    \put(0,0){\includegraphics[width=\unitlength,page=5]{system_model.pdf}}%
    \put(0.36154312,0.05985782){\color[rgb]{0,0,0}\makebox(0,0)[t]{\lineheight{1.36264527}\smash{\begin{tabular}[t]{c}Input Beam\\Pattern\end{tabular}}}}%
    \put(0,0){\includegraphics[width=\unitlength,page=6]{system_model.pdf}}%
    \put(0.62923213,0.31417579){\color[rgb]{0,0,0}\makebox(0,0)[t]{\lineheight{1.36264527}\smash{\begin{tabular}[t]{c}Traffic\\Modeling\end{tabular}}}}%
    \put(0,0){\includegraphics[width=\unitlength,page=7]{system_model.pdf}}%
    \put(0.62923213,0.12329411){\color[rgb]{0,0,0}\makebox(0,0)[t]{\lineheight{1.36264527}\smash{\begin{tabular}[t]{c}Link\\Budgeting\end{tabular}}}}%
    \put(0,0){\includegraphics[width=\unitlength,page=8]{system_model.pdf}}%
    \put(0.89122556,0.31417579){\color[rgb]{0,0,0}\makebox(0,0)[t]{\lineheight{1.36264527}\smash{\begin{tabular}[t]{c}Traffic\\Matrix\end{tabular}}}}%
    \put(0,0){\includegraphics[width=\unitlength,page=9]{system_model.pdf}}%
    \put(0.89122556,0.11955138){\color[rgb]{0,0,0}\makebox(0,0)[t]{\lineheight{1.36264527}\smash{\begin{tabular}[t]{c}Channel\\Matrix\end{tabular}}}}%
    \put(0.89019833,0.42582264){\color[rgb]{0,0,0}\makebox(0,0)[t]{\lineheight{1.36264527}\smash{\begin{tabular}[t]{c}Ouputs\end{tabular}}}}%
    \put(0,0){\includegraphics[width=\unitlength,page=10]{system_model.pdf}}%
  \end{picture}%
\endgroup%